\documentstyle[aps,prd]{revtex}
\begin{document}
\def\bi{\bibitem}
\title{Quantum Gravity Equation In Schroedinger Form 
\\In Minisuperspace Description}
\author{S.Biswas $^{*a),b)}$, A.Shaw $^{**a)}$, B.Modak$^{a)}$ and D.Biswas$^{a)}$ \\
a) Department of Physics, University of Kalyani, West Bengal, India, Pin.- 741235 \\
b) IUCAA, Post bag 4, Ganeshkhind, Pune 411 007, India\\
$*$ email: sbiswas@klyuniv.ernet.in\\
$**$ email:amita@klyuniv.ernet.in}
\date{today}
\maketitle
\smallskip
\begin{abstract}
We start from classical Hamiltonian constraint of general relativity to obtain the
Einstein-Hamiltonian-Jacobi equation. We obtain a time parameter prescription 
demanding that geometry itself determines
the time, not the matter field, such that the time so defined being equivalent
to the time that enters into the Schroedinger equation. Without any reference 
to the Wheeler-DeWitt equation
and without invoking the expansion of exponent in WKB wavefunction in powers of
Planck mass, we obtain an equation for quantum gravity in Schroedinger form containing 
time. We restrict ourselves to a minisuperspace description. Unlike matter 
field equation our equation is equivalent to the Wheeler-DeWitt
equation in the sense that our solutions reproduce also the wavefunction of the
Wheeler-DeWitt equation provided one evaluates the normalization constant
according to the wormhole dominance proposal recently proposed by us.
\end{abstract}
\section{\bf{Introduction}}
\par
There have been various attempts towards a resolution of the problem of time
in quantum general relativity 
\cite{hal:prd,kie:cqg1,kie:prd1,kie:prd2,kie:gr-qc,wad:np,singh:ann,pad:prd,pad:pram}. 
The classical Hamiltonian constraint, in
quantum theory of gravity, leads in an appropriate operator version to the 
Wheeler-DeWitt equation 
\begin{equation}
\hat{H} \Psi =0,
\end{equation}
where $\Psi$ is the wavefunction of the universe. The equation (1) when compared
to a Schroedinger equation does not show the presence of time variable. In short,
this is the problem of time in quantum gravity. Secondly, the interpretation of 
conservation of probability remain obscure without external time. Towards a 
solution of the problem of time in quantum gravity,  
a time variable $t$ is obtained \cite{kie:ttn,kuch:grra} by performing an 
appropriate canonical transformation in which conjugate
momentum $p_t$ occurs linearly such that
\begin{equation}
H=H_r+p_t=0
\end{equation}
in which $H_r$ is the reduced Hamiltonian with conjugate momentum occurring 
quadratically. Upon quantization this becomes 
\begin{equation}
\hat{H}\Psi=o\Leftrightarrow (\hat{H_r}+{\hbar\over i}
{\partial\over {\partial t}})\Psi=0\,.
\end{equation}
Though this approach has been successful in cylindrical gravitational waves or eternal
blackholes, its general viability remains unclear. In the second approach one
starts with the Wheeler-DeWitt equations and looks at a sensible concept of time.
In this approach the normalization of the wavefunction which in turn requires an 
interpretation of `probabilities' still remains unclear. Recent trend suggests
to consider a solution of (3) in the form 
\cite{kie:rag}
\par
\begin{equation}
\Psi({\it{G}},\phi)\simeq C({\it{G}}) 
{\exp \left[({i\over \hbar}) S_o({\it{G}}) \right]} \psi({\it{G}},\phi) 
\end{equation}
where ${\it{G}}$ denotes the gravitational fields, $\phi$ stands for nongravitational
fields and $C$ is a slowing varying prefactor. One then obtains with the 
identification 
\begin{equation}
i\hbar\nabla S_o \nabla \psi\equiv i\hbar {\partial\psi\over {\partial t}}\simeq H_m\psi\,.
\end{equation}
In deriving (5), one uses Wheeler-DeWitt equation with the WKB ansatz (4), but
this has some limitations. The Wheeler-DeWitt equation (3) is linear in $\psi$ and hence
if $\vert\psi_1>$ and $\vert \psi_2>$ are solutions of (3), then the superposition 
principle demands 
$\vert\psi>=a\vert\psi_1>+b\vert\psi_2>$ to be also a solution. There is no a
priori reason why the universe could not be in this state. But experience 
dictates that the universe behaves almost classically as far as macroscopic 
observations are concerned. For such a superposed states e.g., like (4), the 
derivation of (5) cannot be carried out. This is an inherent difficulty if one 
starts with the Wheeler-DeWitt equation. In the present paper we try to resolve
the problem of time using a semiclassical point of view, avoiding the steps (1),
(4) and (5) but keeping the fruits that these equations convey.
\par
In section II we review critically the problem of time and state clearly
the approach that we follow in the present paper. A good non-technical 
review \cite{butt:gr-qc} as well as a technical review \cite{ish:gr-qc} may be
helpful to understand the many points followed in the present work. In
section III we consider a model and solve the classical Hamiltonian constraint 
to obtain the 
prescription of time. This section deals with the emergence of complex paths 
for a model with gravity plus a minimally coupled scalar field in a FRW universe. 
In section IV we obtain the Schroedinger-Wheeler-DeWitt (SWD)
equation for the model described in III. In section V we obtain the solution 
of the SWD equation including the basic aspect of the wormhole dominance
proposal \cite{bis:prd}. Assuming a Gaussian ansatz for SWD solution in the 
Lorentzian region, we show that it gives back the Hartle-Hawking wavefunction 
when continued to Euclidean spacetime. In section VI we make a concluding 
discussion. 
\section{\bf{Time in Quantum Gravity}}
The approach that we follow in the present work is termed as the emergence
of time before quantization, though less persued in the literature. We therefore
spend some words following the ref.\cite{butt:gr-qc,ish:gr-qc}. In quantum gravity dealing with the wavefunction of the universe, the 
traditional Copenhagen interpretation requires an `observer' to carry out 
measurement; unfortunately  we do not have this observer in quantum cosmology. A refinement 
replaces the `observer' by a `classical background' external to the system. This 
`precondition of unambiguous communication', in Bohr's words, is not a well-placed 
argument.
\par
Let us turn to the Everettian idea. According to this idea the particular classical realm, we observe or 
we live in, is basically just `one component' of the universal state vector which 
always evolves deterministically, never collapsing. The criticism against the 
``preferred basis'', chosen to serve as approximate position eigenstates of 
measurement, is now circumvented through the concept of decoherence. The 
decoherence allows a dynamically motivated specification of the `preferred basis'. 
However, there are some subtle aspects: avoidance of the traditional 
no-hidden-variable theorems i.e., algebraic theorems in the tradition of Von 
Neumann, and the non-locality theorems in the tradition of Bell that must be 
avoided in the Everettian `decoherence' mechanism. The decoherence is basically 
a diffusion of coherence (characteristic of quantum superposition principle) 
from a system to its environment. In the dancing coherent-ground of the `system' 
plus the `environment', macro-objects' initial superposition diffuses to the 
environment such that a `variable' (position in ordinary quantum theory, scale 
factor in gravity) emerges nearest in sense to the classical counterpart.
The above discussion refers to any system that deals with quantum to classical 
transition. For systems whose dynamics are described by any of the three 
fundamental interactions (electromagnetic, weak and strong) other than gravity, 
we have a time parameter with three characteristics. In the above three interactions the `time' i.e., 
the time of everyday reality is considered (i) classical, (ii) non-dynamical and
(iii) being the same in all models. But in quantum gravity the situation is 
rather obscure. We have here two broad problems: problem of time and the 
emergence of time. In classical general relativity the condition (i) is 
satisfied, whereas the conditions (ii) and (iii) are not so because the time enters into 
the Einstein equations dynamically just like other three position co-ordinates 
and different geometries (we understand here as different models) evolve with 
different facets of time. Thus for a given manifold with a spacetime structure 
we observe time as if having many fingures - called `many fingured' time. This 
is briefly the problem of time. The problem with the `emergence of time' is to 
embed/graft/bury an approximate physical time in timeless formulation of quantum 
gravity. By `approximate' we intend to satisfy the conditions (i), (ii) and (iii), 
at least approximately.
\par 
At present we have two ways of embedding the time in quantum gravity formulation 
and it is carried out through canonical quantization. The two ways of embedding
the time are to prescribe it `before quantization' or to do the same `after 
quantization'.  Let us discuss the two procedures briefly. In quantum mechanics 
we follow the `constrained quantization' in which the Hamilton of the system 
contains more variables than the physical degrees of freedom. For example, the 
classical general relativity constraint $H=0$ is converted to a constrained 
quantum equation by replacing $p_i=-i{\partial\over {\partial q_i}}$.
After that one introduces the time by a prescription. This is known as `time 
after quantization'. 
Another attempt (of course less perused) is to solve the constraints before 
quantizing and to find the so called `internal time' as a function of the 
canonical variables of general relativity such that this `time' could serve as a 
time for the Schroedinger  equation of quantized theory. This procedure is named 
as `time before quantization'. We follow the second one in the present paper.   
\par
It is worthwhile to point out the reasons for choosing the embedding before 
quantization. In the approach `before quantization', the constrained 
quantum equation is the Wheeler-DeWitt equation $\hat{H}\Psi=0$ and obviously 
this is timeless in character. We should have a way to 
interpret the wave function in a timeless way i.e., one has to settle between the 
`Copenhagen interpretation' and the `Everettian interpretation' that includes now 
the decoherence mechanism compared to the collapse of wavefunction of the former 
interpretation. This is the first problem. The second problems relates to the 
initial conditions for the Wheeler-DeWitt wavefunction, now classed by the words 
`wavefunction debate' because of various boundary condition proposals; namely, 
Hartle-Hawking no boundary proposal \cite{har:prd}, Vilenkin's tunneling 
proposal \cite{vil:prd} and Linde's proposals \cite{lin:jetp}. Recently 
another proposal \cite{bis:prd} has been made considering complex solutions of Hamilton-Jacobi 
equation of classical general relativity. In the proposal, termed as `wormhole 
dominance proposal' attempts have been 
made to include complex solutions in the framework of the Wheeler-DeWitt 
equation with a general prescription of retaining `allowable' and/or `good' 
complex WKB paths. The `wormhole dominance proposal' serves as a naive 
attempt.There were many attempts \cite{hall:prd2,lyo:prd,hall:prd3} 
to consider complex solution while evaluating the wave function of the
universe using the path integral formulation but the recent attempts 
\cite{unru:gr-qc}
suit nicely to the wormhole dominance proposal. The choosing of
the paths in WKB formulation or in the Hartle-Hawking path integral
formulation  is another problem and needs to incorporate
complex solutions/complex paths. This is the second problem. In 
the present work we do not delve into the controversy while keeping only a 
salient feature in evaluating the normalization constant of Schroedinger-Wheeler
-DeWitt solution. The third problem is related to the association of the 
wavefunction to the Lorentzian condition in quantum gravity.
\par
In the past the approach to quantum cosmology considered the behaviour of the wavefunction, 
as a function of the overall scalefactor, $a$ of the 3-metric $h_{ij}$, on the 
spacelike surface $\sum$. If the dependence on $a$ is exponential, the 
wavefunction corresponds to the Euclidean spacetime, while an oscillatory 
dependence on $a$ is interpreted as corresponding to a Lorentzian spacetime. 
However this distinction between exponential and oscillatory is not precise, 
and does not identify, which part of the wavefunction describes which physical 
situation \cite{bou:prd}. Technically it amounts to asking whether a given 
spacelike surface is a part of a Lorentzian spacetime or a Euclidean spacetime. Our approach 
is: avoid the Wheeler-DeWitt equation, solve the classical constraints equations 
with a prescription for embedding the time, put the constraint equation in a form 
relating $p_t$ and $p_\phi$ (identifying $p_t$ and $p_\phi$ from Hamilton-Jacobi 
equation), quantize the constraint equation (we call it Schroedinger-Wheeler-
DeWitt equation), adopt a boundary condition for the wavefunction with a view to 
decoherence, continue the wavefunction to the Euclidean regime and see whether 
the boundary conditions of classical spacetime (i.e., a large region) is any 
how related to the all quantum region (i.e., where both gravity and matter field are 
quantized). This approach allows to avoid many of the problems that have already 
been mentioned above. This is the main content of the present paper.
\section{\bf{Model}}
We start with an action
\begin{eqnarray}
I & = & {\int {{d^4} x {\sqrt{-g}}\left[ {-R \over {16 \pi G}} - 
{1 \over {2 {\pi}^2}} ({1 \over 2} \phi_{,\, \mu} \phi^{,\,\mu} 
+ V(\phi) ) \right] } } \nonumber \\
& - & {1 \over {8 \pi G}}{\int_{\sum} d^3 x \sqrt{h} K}
\end{eqnarray}
in a FRW universe
\begin{equation}
ds^2 = - dt^2 + a^2 (t) \left[ {{{dr^2} \over {1 - k r^2}} + r^2 (d {\theta}^2
+{\sin}^2 \theta d {\phi}^2)} \right].
\end{equation}
The Hamiltonian constraint corresponding to (6) now reads
\begin{equation}
H=-{1 \over {2Ma}} {P_a}^2 + {1 \over {2 a^3}}{P_{\phi}}^2 - {M \over 2} ka 
+ a^3 V(\phi) = 0\,.
\end{equation}
In equation (8), $M =  {{3 \pi}\over {2G}}$, where $G$ is Newton's constant,  
$k  =  0, \pm 1 $ for flat, closed and open models, and
$P_a = - M a {\dot{a}},\; P_{\phi} = a^3 {\dot{\phi}}\,.$  
Identifying
$P_i = {{\partial S} \over {\partial q_i}},\; q_i=a,\; \phi,$ the 
Einstein-Hamilton-Jacobi equation is 
\begin{equation}
{-{ 1 \over {2M}} }{({{\partial S} \over {\partial a}})^2} + {1 \over {2a^2}}
{({{\partial S} \over {\partial \phi}})^2} -{1 \over 2}Mka^2 + a^4 V(\phi) = 0\,.
\end{equation}
Now to obtain the prescription like (5), we define a time operator consistent with
Liouville's equation as
\begin{eqnarray}
{\partial\over {\partial t}}&=& \sum_{i}({{{\partial H}\over {\partial P_i}}
{\partial\over {\partial q_i}}-{{\partial H}\over {\partial q_i}}
{\partial\over {\partial P_i}}})\nonumber \\
&=&
{{\partial H}\over {\partial P_a}} {\partial\over {\partial a}}
+{{\partial H}\over {\partial P_\phi}} {\partial\over {\partial \phi}}
-{{\partial H}\over {\partial a}} {\partial\over {\partial P_a}}
-{{\partial H}\over {\partial \phi}} {\partial\over {\partial P_\phi}}\,.
\end{eqnarray}
that would satisfy (8) or (9).
From (10) we get
\begin{eqnarray}
{{\partial a}\over {\partial t}}&=& {{\partial H}\over {\partial P_a}}=
-{P_a\over Ma}\\
{{\partial \phi}\over {\partial t}}&=&{ 1\over a^3} P_\phi\\
{{\partial P_a}\over{\partial t}}&=& -[{{P_a}^2\over {2Ma^2}}
-{3{P_\phi}^2\over {2a^4}}-{Mk\over 2}+3a^2V(\phi)]\\
{\partial P_\phi}\over {\partial t}&=&-a^3{{\partial V}\over{\partial \phi}}
\end{eqnarray}
The above four equations along with the Hamiltonian constraint (8) determine
t and also the paths characterized by $a$. Our aim is to find the extrema of the
action when $ a,t~~  and~~  \phi $ are complex i.e., we are looking at complex 
four-metrics and complex fields on a real four dimensional manifold characterized
by the coordinates $ t, r, \theta , \phi $. To fix the origin of time defined above
and to impose regularity condition on the four geometry at $ a=0 $, we choose
the boundary condition as \cite{unru:gr-qc}
\begin{equation}
a(t=0)=0,~~ {{\partial a}\over{ \partial t}}= \beta ~i~ at~ t=0,
\end{equation}
where $\beta=\pm 1$.
For analytical simplicity we also assume that $V(\phi)$ is approximately constant,
near about the region where (15) is satisfied.
We find using (12) and (14)
\begin{equation}
a^3 {{\partial \phi}\over {\partial t}}= constant.
\end{equation}
If this constant is non-zero the boundary condition at $ t=0 $ leads
$ {{\partial \phi}\over {\partial t}}\rightarrow \infty $. Hence we should
have $ \phi=constant $. This implies $ P_\phi=0 $. Under this condition,
when $V(\phi)\simeq V_0 $ the solutions of (11)- (14) are called the
zeroth order solutions. We write 
\begin{equation}
P_a\approx {P_a}^0={{\partial S_0}\over {\partial a}}
\end{equation}
where $ S_0= S(a,V_0) $ is now identified as zeroth order action.
\par
Restricting ourselves to k=+1 universe and using the (13) and the constraint
equation (9) we get 
\begin{equation}
 a'^2 + aa'' =4a^2 {V_0\over M} -1,
\end{equation}
where the prime now denotes differentiation with respect to $t$.
Denoting its solution by $ a_0(t) $ we get
\begin{eqnarray}
a_0(t)&=& \beta{\sqrt{M\over 2V_0}} sin{(i\sqrt{2V_0\over M}t)},\nonumber\\
    &=&{ \frac {\beta} {\nu}} \sin{(i~\nu~t)},
\end{eqnarray}
where $\nu=\sqrt{2V_0/M}$.
This solution satisfies the boundary conditions
\begin{equation}
a(0)=0,~~ a'(0)=\beta~ i=constant,~~ a''(0)=0
\end{equation}
\par
Let us construct the zeroth order action $ S_0 $ for the solution (19).
Using (11) and (19) we find
\begin{equation}
P_a^0= {\frac {\partial S_0} {\partial a}}= -M\beta a_0 
(2a^2\frac {V_0} {M}-1)^{1/2}.
\end{equation}
We note that for $2a^2\frac {V_0}{M}>1$, $P_a^0$ is real and is identified as
classically allowed region. For $2a^2\frac {V_0}{M}<1,~P_a^0$ is imaginary
and the corresponding region is identified as classically unallowed region.
Thus we have the turning points at $a_0=0 $ and at $a_T=\sqrt{\frac {M}{2V_0}}$.  
We put the final boundary condition as $a_0 =0$ at $t=0,$ and $a_0=a_f$ at $t=t_f$.
Integrating (21) we get
\begin{equation}
S_0(a_f, V_0)= \frac {-M^2\beta} {6V_0} (2a^2\frac {V_0} {M}-1)^{3/2} + i~\frac {M^2} {6V_0}
\end{equation}
We have used $ (-1)^{3/2} = \beta~~i $. The two possible values of the action
(due to the presence of $\beta$)
have been the subject of immense controversy. However, we will not dwell 
upon this controversy.
Recently there has also been some controversy with the `factor ordering' problem
\cite{kon:prd} 
when one tries to construct the wavefunction from the Wheeler-DeWitt equation's
solution with a given boundary condition proposal. The wormhole
dominance proposal proposed by one of the authors \cite{bis:prd} dwells upon 
complex path
approach in the WKB approximation. The present work, we find, substantiate our 
previous
work and confirm the prescription given in the wormhole dominance proposal. In 
the present work we discuss briefly the way how the complex paths enter into
the description of wavefunction calculation of the Schroedinger-Wheeler-DeWitt 
equation. The details will be placed elsewhere.
Here we mention the salient results that will be needed in the present discussion.
\par
For $V(\phi)$ varying slowly such that $V(\phi)-V_0<<V_0$ we calculated the corrections $\delta S,~~\delta^2 S $ finding the first order
correction to $a_0$ and $\phi_0$. From the calculation we observe that 
the semiclassical solution of Einstein equation allows complex solutions in both 
$a$ and $t$ space. The Hamilton-Jacobi function has not only real solutions but 
complex solutions that also contribute to the semiclassical path integral while 
evaluating the wavefunction of universe according to 
\begin{equation}
\Psi (G^{(3)},\Phi (x\in G^{(3)}))=\int{e^{iS(G^{(4)},\phi(x\in G^{(4)}))}\delta \phi\delta G}.
\end{equation}
Here $G^{(4)}$ represents the four geometry and the fields $\phi$ defined on 
this four geometry are non-singular. There are some technical difficulties in 
choosing the complex paths in $a$-space in (23). 
However no unique prescription still has emerged. The question is to choose or find 
`allowable' and `good' paths, 
that contribute in (23). Finally, once one has calculated the wavefunction, it
is still unclear how it resolves the problem of time, settles between the 
Copenhagen and Everettian approaches and interpret the timeless version of the 
Euclidean path integration. The wormhole dominance proposal provides a prescription 
for that in evaluating $\psi$ within the framework of complex semiclassical WKB 
approximation. A more technical discussion relating the complex paths and the 
evaluation of Euclidean path integral is reserved for concluding discussion in 
order to not track away from the main content of the paper.
Let us try to understand the complex solutions considering (19) and (22).
An important boundary condition is that both the initial and final value 
of $a$ must be real, keeping option open to include the complex paths in 
evaluating (23). Let us investigate the nature of the paths. We have
$a(t_i=0)=0 $ and take $a(t=t_f)=a_f$.  
\par
For a given $a_f=real,$ we find from (18)
\begin{equation}
\beta \cosh{(\nu t_R)}\sin{(\nu t_I)}=\nu a_f
\end{equation}
\begin{equation}
\cos{(\nu t_I)}=0
\end{equation}
where $t_f=t_R+i~t_I$. Corresponding to different $t_i$ given by (25) we have the 
same $t_R$ for a given real $a_f$. In otherwords, there are many different end 
points $t_f$ which give the same $a_f$ and many different contours to each of 
these end points. A numerical plot of the paths in the complex scale factor $(a)$
space of solution for various complex time $t$ will be found in \cite{unru:gr-qc}.
In the cited work, they deal with complex solutions to study the classical 
evoluation without showing the construction of the wavefunction but in our work 
we construct the wavefunction taking the contribution of complex paths thereby 
exposing the connection between the small $a$ and large $a$ boundary conditions 
on the wavefunction.
\par
Let us calculate the zeroth order action
\begin{eqnarray}
S_0(a_f,\,\phi_f) &=& \int^{t_0}_{0}P_{a}^{0} \,a'_0\, dt\nonumber \\
&=& -M\int^{t_0}_{0} a_0(a'_0)^2\,dt\nonumber \\
&=& -M\int^{a_f}_{0} a_0 a'_0\,da_0
\end{eqnarray}
and consider
\begin{equation}
\Psi_0\,( a_f,\,\phi_f)\propto e^{iS_0},
\end{equation}
as the wavefunction of the universe (assuming that this is possible in the approach we follow).
According to the wormhole dominance proposal, the wavefunction $\Psi$ given in 
(27) must be multiplied by those contributions of the contours in complex 
$a$-space (i.e., in complex $t$ space) which have the end points $a_f$ and 
initial point $a(0)=0$. Looking at (21), we find 
\begin{equation}
P^{0}_{a}={{\partial S_0}\over {\partial a}}=-Ma_0(2a^{2}_{0}{V_0\over M}-1)^{1/2}
\end{equation}
which have turning points at $a_0=0$ and $a_0=\sqrt{{M}\over {2V_0}}=a_T$. Thus,
\begin{equation}
S_0 (Complex\;\;paths\,)=S_0(a_f,\,0)+S(a_0=0\rightarrow a_T \rightarrow a_0
\rightarrow a_f).
\end{equation}
Here the first term is the contribution from the path that goes from $a=0$ to
$a=a_f$ and the second term when evaluated corresponds to a loop path between 
$a=0$ and $a_f=\sqrt{\frac {2V_0} {M}}$ times the first term in (29). When the 
repeated loops contribute we get a geometric series. Correspondingly the
wavefunction is given by
\begin{equation}
\Psi(a_f,\,\phi_f)={{e^{iS_0(a_T,0)}}\over
{1-e^{2iS_0(a_T,0)}}}\,\Psi_0,
\end{equation}
considering repeated paths, to and fro, between $a=0$ and $a=a_T$. For
the derivation of this result the reader is referred to \cite{bis:prd}.
The contribution $S_0(a_T,0)$ gives the second term in (21) and we observe that
the imaginary part of the action determines the amplitude and the real part determines
the phase of the wavefunction. The CWKB approach gives the extra term i.e., 
the denominator in (30). This term correctly takes into account the one loop
quantum correction. Applying this technique to particle production in uniform 
electric field, we obtained \cite{bis:cqg} the Schwinger's one loop result 
\cite{sch:prd}
of $ Im\,L_{eff}^{(1)}$ term by term and observe that the particle production
occurs due to the instability in the Hartle-Hawking vacuum. This lends 
support to keep the denominator in (30) which some authors (see concluding
discussion) feel not necessary in the evaluation of the path integral.
\par
To apply this approach in our case, we need a time contained description
that will render a physical interpretation of $(1-exp(2i\,S))^{-1} $ term.
Our approach is to treat
the gravity as classical and retaining for $\phi$ the quantum behaviour and to 
obtain an quantum equation that contains the time.
\section{\bf{Schroedinger-Wheeler-DeWitt Equation}}
In our approach we keep only the $\delta\,S\equiv S_1$ term and consider the 
region in which $\nu\,a_f\gg 1$ and the gravity  field could be treated 
classically such that $S_1<< S_0$ in the region of interest.
By this we allow $a$ to be complex and is given by (19). We now eliminate
$S_1$ through the following steps.
In the region $\nu a_f>>1$ we take (see (11))  
\begin{equation}
{\partial \over {\partial t}}=-{1\over {Ma}}{{\partial S_o}\over {\partial a}} 
{{\partial}\over {\partial a}}\,.
\end{equation}
At present we omit the subscript $0$ in $a_0$ and will be introduced when 
required.
Equation (31) now gives the prescription of time. The $\partial\over {\partial t}$ 
is a directional derivative along each of the classical spacetimes which can be viewed
as classical `trajectories' in the gravitational configuration space. In view 
of (11)-(13), we identify the variable $t$ as the classical time parameter. Using
(31), we now write
\begin{equation}
{{\partial S} \over {\partial t}}=-{1\over {Ma}}{{\partial S_o}\over {\partial a}}
{{\partial S}\over {\partial a}}\,,
\end{equation}
where $S$ is given in (9).
Writing $S(a,\phi)=S_o(a)+S_1(a,\phi)$, we write (32) as
\begin{equation}
{{\partial S} \over {\partial t}}=-{1\over {Ma}}({{\partial S}\over {\partial a}})^2
+{1\over {Ma}}{{\partial S_1}\over {\partial a}}{{\partial S}\over {\partial a}}\,.
\end{equation}
We now substitute $({{\partial S}\over {\partial a}})^2$ from (9) to get
\begin{equation}
{{\partial S} \over {\partial t}}=-{1\over {a^3}}({{\partial S}\over {\partial \phi}})^2
+Mka-2a^3V(\phi)
+{1\over {Ma}}{{\partial S_1}\over {\partial a}}{{\partial S}\over {\partial a}}\,.
\end{equation}
In the region of interest i.e., when $V(\phi)\approx V_0$,
$P_\phi^0=0$,  we write $P_a\approx P_a^0$ and using (13) we find
\begin{equation}
\frac {1}{Ma}(P_a^0)^2=2a[(-P_a^0)'+\frac {Mk}{2}-3a^2V_0].
\end{equation}
Further neglecting $(\frac {\partial S_1}{\partial a})^2 $ term we get
\begin{eqnarray}
\frac {1}{Ma}\frac {\partial S_1}{\partial a}\frac {\partial S}{\partial a}&\simeq&
\frac {1}{Ma}\frac {\partial S_1}{\partial a}\frac {\partial S_0}{\partial a}\nonumber\\&=&
\frac {1}{Ma}\frac {\partial S_0}{\partial a}\frac {\partial S}{\partial a} -
\frac {1}{Ma}(\frac {\partial S_0}{\partial a})^2 \nonumber \\&=&
\frac {1}{Ma}\frac {\partial S_0}{\partial a}\frac {\partial S}{\partial a}-
\frac {1}{Ma}(P_a^0)^2.
\end{eqnarray}
Using (32) to replace the first term in (36) by $-\frac {\partial S}{\partial t}$
and using (35) to replace the second term in (36) we get from (34)
\begin{equation}
\frac {\partial S}{\partial t}=-\frac {1}{2a^3}(\frac {\partial S}{\partial
\phi})^2
-a^3 V(\phi)+a(P_a^0)'+3a^3 V_0.  
\end{equation}
When $V(\phi)\approx V_0$, the Hamiltonian constraint is
\begin{equation}
-\frac {(P_a^0)^2}{2Ma} - \frac {Mka}{2} + a^3 V_0=0
\end{equation}
Using (38) in (35) we get
\begin{equation}
(P_a^0)'= Mk - 4a^2 V_0
\end{equation}
In the region, $4a^2\frac{V_0}{M}>> 1$, (we now consider $k=+1$) we have
\begin{equation}
a(P_a^0)'=-4a^3V_0
\end{equation}
Putting this value in (37), we finally get 
\begin{equation}
\frac {\partial S}{\partial t}= -\frac {1}{2a^3}(\frac {\partial S}{\partial \phi})^2
-a^3(V(\phi)+V_0).
\end{equation}
It is now straightforward to quantize (41) identifying 
$P_t={{\partial S}\over {\partial t}}$ and
$P_\phi=\frac {\partial S}{\partial \phi}$. Upon quantization we get
\begin{equation}
i \frac {\partial \Psi(a,\,\phi)}{\partial t}=[-{\frac {1}{2a^3}}
\frac {\partial^2}{\partial\phi^2} +a^3(V(\phi)+V_0)]\Psi(a,\,\phi).
\end{equation} 
In (42), $\Psi$ refers to the wavefunction of the universe, in which $a(t)$ is 
given by the solution of classical constraint equation. In this sense, apart 
from the extra term $V_0$ (that one does not have through `time' after 
quantization) (42) takes the equivalent role as does the Wheeler-DeWitt equation. 
We will show this in the next section.
In obtaining (42) we have nowhere used the Wheeler-DeWitt equation and the 
corresponding WKB ansatz for $\Psi$. Equation (42) is our Schroedinger-Wheeler-DeWitt
equation. We need some boundary condition for the SWD wavefunction. We would 
start with a Gaussian ansatz and see what type of wavefunction does it lead to when 
continued to a region $2a^2\frac{V_0} {M}<1$, where both gravity and the matter 
field are both quantum in nature.
From the classical Einstein equation
\begin {equation}
G_{\mu\nu}=R_{\mu\nu}-{1\over 2} g_{\mu\nu} R=kT_{\mu\nu}\,
\end{equation}
we observe that ``geometry and matter'' get coupled through (43). It is also a 
well known fact that the matter field is quantized and for that reason in equation 
(43) one writes $<T_{\mu\nu}>$ on the right hand side and treats $g_{\mu\nu}$ as
classical background. Keeping this in mind we argued that time is determined by the 
geometry itself.
Let us be precise about the equation (42). The time that appears in (42)
is the time that we use in quantum theory and is specified by the
classical background $a(t)$ which in turn has been determined from the
initial condition as if the many fingureness gets hidden in the $a(t)$
by the initial condition. The dynamicalness of $t$ has been transferred
by demanding both $a(t)$ and $t$ complex and allowing $t$ to move both
forward and backward direction. Whereas it is difficult to realize it in
the framework of timeless Wheeler-DeWitt equation, but this
interpretation works for equation (42). What is important to show that
whether this time is buried in the Wheeler-DeWitt equation or not. We
will take up this point in the next section.
\section{\bf{Solution of Schroedinger Wheeler-Dewitt Equation}}
Let us now take the solution of (42)
\begin{equation}  
\Psi = N (t) e^{- {{\Omega (t)} \over 2} \phi^2},
\end{equation} 
with a choice of $V(\phi)={\lambda\over 2}(1-m^2\phi^2), \lambda$ being a constant. The time 
appearing in (42) is the physical time, at least in the region we are considering.
Now substituting (44) in (42) we find 
\begin{equation}
i{d \over {dt}} \ln N = {\Omega \over {2a^3}} + {a^3 \lambda},  
\end{equation}
\begin{equation}
i{ {\partial\Omega}\over {\partial t}} = {{\Omega^2 + a^6 \lambda m^2} \over
a^3}\,.
\end{equation}
Making an substitution 
\begin{equation}
\Omega = - i a^3 {{\dot{y}} \over y},
\end{equation}
and using the conformal coordinate $dt=ad\eta$, equation (46) gives
\begin{equation}
y'' + 2 {{a'} \over a} y' - \lambda m^2 a^2 y = 0\,.
\end{equation}
Taking an inflationary background $a(\eta)=-{1\over
{\sqrt{\lambda}\eta}}$,
the solution $y$ is given by 
\begin{equation}
y =C_1 \eta^{{3/2} \pm {\sqrt{{9/4} + m^2}}}\,.
\end{equation}
For practical purposes 
$m^2<{9\over 4}\, and\,\sqrt{{9/4} + m^2} \simeq ({3/2} +{{m^2}/3}).$
Taking the negative sign in (49) and in conformal coordinate
\begin{equation}
\Omega=-ia^2(\eta){y'\over y}\,,
\end{equation}
we get
\begin{equation}
\Omega = -i m^2 \sqrt{\lambda} {a^3 \over 3}.
\end{equation}
Since $m^2\sqrt{\lambda}$ is small and $a$ is large and 
${d\over {dt}}=\sqrt{\lambda}a{d\over {da}}$, we get from (45)
\begin{equation}
N = N_o \exp { \left[ - {{i a^3 \sqrt{\lambda}} \over 3} \right]},
\end{equation}
the integration constant $N_o$ has to be determined through a suitable 
initial condition.We now write down the solution of the SWD equation
(42) as
\begin{eqnarray}
\Psi &=& N_0\exp{\left[\,i\frac{ m^2\sqrt{\lambda} a^3}{6}\phi^2
-i\frac {a^3\sqrt{\lambda}}{3}\right]} \nonumber \\
&\simeq& N_0\exp{\left[\,-i\frac
{a^3[\lambda(1-m^2\phi^2)]^{1/2}}{3}\,\right]}\nonumber\\
     &\simeq& N_0\exp{\left[\,-i~\frac{(a^2 2V)^{3/2}}{3.2V}\,\right]}\nonumber
     \\
     &\simeq& N_0\exp{\left[\,-i~\frac{(a^2 2V-1)^{3/2}}{3.2V}\,\right]}
\end{eqnarray}
In obtaining (53) we have made the approximation
$\sqrt{\lambda}(1-{1\over 2}m^2\phi^2)
\simeq (1-m^2\phi^2)^{1\over 2}\sqrt{\lambda}=\sqrt{2V(\phi)}$  
and $2a^2V\simeq 2a^2V-1$ for $2a^2V>1$, which is quite valid in
classically allowed region. This is what we meant earlier by saying `the
region
of interest' while neglecting $S_1$ compared to $S_o$. Obviously the 
wavefunction (53) is not normalizable because of absence of real part of
$\Omega$
in equation (51). The general procedure is to consider higher mode solutions
of
the scalar field. We take however the complex trajectory approach as in 
\cite{bis:prd}. It should be pointed out that though we started with a Gaussian 
form, the wavefunction very closely resembles the WKB form that we guessed 
earlier (see (27)). To compare our result with that of the Wheeler-DeWitt 
equation, let us obtain the Wheeler-DeWitt equation corresponding to the above 
potential. We pass over a M-independent description with the following 
substitution in the Hamilton-Jacobi equation (9).
\begin{equation}
S\rightarrow M~S,~~ \phi\rightarrow M{\phi}^{\frac {1}{2}},
~~V(\phi)\rightarrow M~V(\phi).
\end{equation}
The Wheeler-DeWitt equation that follows from (8) or (9) is
\begin{equation}
\left[{{\partial^2}\over {\partial a^2}}-{1\over a^2}{{\partial^2}\over
{\partial \phi^2}}-a^2(1-a^22V(\phi))\right]\Psi=0.
\end{equation}
As is evident from the classical equation (28) or the Wheeler-DeWitt equation 
(55), we find the two turning points at $a_0=0$ and at 
$a_{\sc{T}}={1\over {\sqrt{2V}}}$. It should be pointed out that we have 
confined the discussions to regions in which the potential $V(\phi)$ can be 
approximated by a constant $V_0$, acting like a cosmological constant, so that 
the $\phi$ dependence in (55) can be effectively ignored. We now write (53) as
\begin{equation}
\Psi=N_0\,e^{-iS_{eff}}
\end{equation}
where
\begin{equation}
S_{eff}={{(a^22V-1)^{3/2}}\over {3\times 2V}}.
\end{equation}
According to `wormhole dominance' proposal, $N_0$ is given by 
\begin{equation}
N_o = {{\exp \left[{-iS_{eff} (a_{\sc{T}},0)}\right]} \over 
{1 - \exp {\left[- 2i S_{eff} (a_{\sc{T}},0) \right]}} }\,,
\end{equation}
where
\begin{equation}
S_{eff} (a_{\sc{T}}, 0) = S_{eff}{\vert_{0}^{a_{\sc{T}}}}\,. 
\end{equation}
Evaluating (59) we find
\begin{equation}
N_o = {{\exp ({1\over {3\times 2V}})} \over 
{(1 - e^{2\over {3\times 2V}})}}\,
\end{equation}
taking $(-1)^{3/2}=-i$. Continuing in the region $2a^2V\,<\,1$, we get 
\begin{equation}
\Psi=C_1e^{{1\over {3\times 2V}}\left[1-(1-a^22V)^{3\over 2}\right]}\,,
\end{equation}
where
\begin{equation}
C_1={1\over {1-\exp({2\over {3\times 2V}})}}\,.
\end{equation}
Equation (61) is the wavefunction of wormhole-dominance proposal. If we leave
aside the $C_1$ term in (61), it gives the Hartle-Hawking wavefunction when 
$V(\phi)=Constant\;=V$. In our earlier work \cite{bis:prd} 
we have shown that the normalization constant $N_o$  
thus obtained is consistent with Coleman and Klebanov's 
\cite{col:np,kle:np} arguments and 
we interpreted the constant $N_o$ as a manifestation of wormhole contribution
to the wavefunction.The recent work \cite{unru:gr-qc} also supports our 
viewpoint.
\par
Thus we have shown that the initial condition of SWD equation is fixed by the 
Gaussian anstaz
\begin{equation}
\Psi = N (t) e^{- {{\Omega} \over 2} \phi^2}
\end{equation}
in which a factor $N(t)$ is obtained taking the contributions of 
repeated reflections between the turning points. This contribution can be 
interpreted as due to wormhole contributions.
\par
The superposition principle of quantum mechanics requires the existence of repeated
reflections and here it is ensured by the presence of turning points. The classical
turning point $a={1\over \sqrt{2V}}$ serves as a doorway to keep the universe's
emergence in the classical regime. The quantum turning point $a=0$ acts as an
entrance door of all quantum force supposed to arise from the other ensemble 
of universes.
\par
An intuitive picture is that there is an ensemble of `Quantum Universes' from
which our universe tunnels retaining only the quantum property of uncertainty
principle, the other quantum characteristics of ensemble of universe are thus
still lying hidden and of course this applies to our universe also. 
\par
\section{\bf{Discussion}}
At present, all three standard proposals mentioned earlier are formulated in terms of a
Euclidean path integral. This path integral depends crucially on how paths
are chosen to evaluate the the integral. The Euclidean action for gravity
is unbounded from below and hence a complex contour is generally necessary
for convergence. No proposal does uniquely fix the contour for the evaluation
of the path integral and as it stands does not define a unique wave
function of the universe. Klebanov, Susskind and Banks \cite{kle:np} 
evaluated the path integral (in the 3+1 case) considering multiple spheres
configurations connected by the wormholes. However, Halliwell and Myers' 
\cite{hall:prd2} analysis reduces the path integral to a single path integration
over the lapse function and the lapse has to be complex for convergence.
Now as we find, the integration over the complex lapse equivalently be 
considered as an integration over the complex time (using some gauge
condition for the lapse). The scale factor is then complex and a function of 
complex $t$. In the path integral formulation
one evaluates the saddle points that represent the classical solutions. The
result then depends on how the contour is chosen to include or exclude the 
saddles i.e., we require a suitable steepest-descent path. In \cite{hall:prd2}
it is argued not to take the contributions of each saddles by summing them
up as in \cite{kle:np}.
\par
In substance, while evaluating the wavefunction of the universe one must 
incorporate the complex solutions (i.e., complex $a(t)$ with complex $t$)
remembering the fact that both the initial and final $a$ must be real. If 
we translate this viewpoint in terms of WKB formulation, this amounts to 
saying that the wavefunction at a real point $a$ is contributed not only
by real WKB trajectories but also gets contribution from complex trajectories.
Such a WKB formulation with complex trajectories was given in 
\cite{kno:anp} with a heuristic expression for the solution of one 
dimensional Schroedinger equation. We used this technique in the wormhole
dominance proposal \cite{bis:prd} in a simple minisuperspace description.
The technique of CWKB (complex path WKB approximation) applied to other fields
(particle production in curved spacetime and in heavy ion scattering) gives
remarkable results. It should be pointed out that the CWKB construction of the wavefunction
of the universe is basically the same in spirit to that of \cite{hal:prd,kle:np}, 
though more technical aspects remained to be investigated.
\par
To elucidate the effectiveness of our proposal \cite{bis:prd}, we derive a Schroedinger type equation     
avoiding the Wheeler-DeWitt equation, the WKB ansatz and also the path 
integral formulation using the approach: time before quantization. With
a Gaussian ansatz (dictated from decoherence) as a boundary condition, we obtained
the normalization constant of our wavefunction using the wormhole dominance
proposal.
The resulting wavefunction seemingly resembles the Hartle-Hawking wavefunction
of no boundary proposal. If we leave aside the $(1 +exp(2iS))^{-1}$ term of 
$N_0$, the wave function results in the same situation as in \cite{hal:prd}.
(Halliwell and Myers worked in $2+1$ whereas we worked in $3+1$ dimensions)
In our work multiple sphere configurations connected by wormholes are 
interpreted in terms of repeated reflections from the turning points. In the present
work we are thus one step further to identifying the real and imaginary part of 
the action being associated with Euclidean and Lorentzian spacetime respectively.
Though a naive attempt, we have been able to show that the time is buried in the
structure of the Wheeler-DeWitt equation, and manifests itself through
the Schroedinger equation dictated by the matter field Hamiltonian. The 
emergence of Hartle-Hawking wavefunction (better to say, a solution of 
the Wheeler-DeWitt equation) from the solution of the Schroedinger 
equation with a Gaussian ansatz is quite surprising and lends support 
to the Everettian idea of decoherence. The Copenhagen probabilistic
interpretations also remain workable in our approach that had already been 
discussed in \cite{bis:prd}. 

\smallskip
\begin{center}
{\bf{Acknowledgment}}
\end{center}
A. Shaw acknowledges the financial support from ICSC World Laboratory, LAUSSANE
during the course of the work.
\end{document}